\documentclass[twocolumn,pra,aps,showpacs]{revtex4}

\usepackage{mathptmx}
\usepackage{subfigure}
\usepackage{psfrag,graphicx}
\usepackage{dcolumn}
\usepackage{amsmath,amssymb}
\usepackage{bm}
\usepackage{color}
\usepackage{latexsym}
\usepackage{epstopdf}
\usepackage{color}
\usepackage[english]{babel}
\usepackage{latexsym}
\usepackage{psfrag,graphicx}
\usepackage{subfigure}
\usepackage{amsmath}
\usepackage{amssymb}
\usepackage{amsfonts}
\usepackage{bm}
\usepackage{natbib}
\usepackage{epstopdf}
\usepackage{appendix}

\definecolor{mygrey}{gray}{0.35}
\definecolor{myblue}{rgb}{0.2,0.2,0.8}
\definecolor{myzard}{cmyk}{0,0,0.05,0}
\definecolor{mywhite}{rgb}{1,1,1}
\definecolor{mywhite}{rgb}{1,1,1}
\definecolor{myred}{rgb}{1,0.,0.3}

\usepackage[colorlinks=true,citecolor=myblue,linkcolor=myred]{hyperref}

\def\ba{\begin{align}}
\def\enda{\end{align}}
\def\bi{\begin{itemize}}
\def\ei{\end{itemize}}

\def\be{\begin{equation}}
\def\ee{\end{equation}}
\def\bea{\begin{eqnarray}}
\def\eea{\end{eqnarray}}
\def\bse{\begin{subequations}}
\def\ese{\end{subequations}}

\def\i{\text{i}}

\begin{document}

\author{Lachezar S. Simeonov}
\affiliation{Department of Physics, St. Kliment Ohridski University of Sofia, 5 James Bourchier blvd, 1164 Sofia, Bulgaria}
\author{Nikolay V. Vitanov}
\affiliation{Department of Physics, St. Kliment Ohridski University of Sofia, 5 James Bourchier blvd, 1164 Sofia, Bulgaria}
\author{Peter A. Ivanov}
\affiliation{Department of Physics, St. Kliment Ohridski University of Sofia, 5 James Bourchier blvd, 1164 Sofia, Bulgaria}

\title{Compensation of the trap-induced quadrupole interaction in trapped Rydberg ions}
\date{\today }

\begin{abstract}
The quadrupole interaction between the Rydberg electronic states of a Rydberg ion and the radio frequency electric field of the ion trap is analyzed. Such a coupling is negligible for the lowest energy levels of a trapped ion but it is important for a trapped Rydberg ion due to its large electric dipole moment. This coupling cannot be neglected by the
standard rotating-wave approximation because it is comparable to the frequency of the trapping electric field. We investigate the effect of the quadrupole coupling by performing a suitable effective representation of the Hamiltonian. For a single ion we show that in this effective picture the quadrupole interaction is replaced by rescaled laser intensities and additional Stark shifts of the Rydberg levels. Hence this detrimental quadrupole coupling can be efficiently compensated by an appropriate increase of the Rabi frequencies. Moreover, we consider the strong dipole-dipole interaction between a pair of Rydberg ions in the presence of the quadrupole coupling. In the effective representation we observe reducing of the dipole-dipole coupling as well as additional spin-spin interaction.
\end{abstract}

%We consider an ion chain of Rydberg ions as well as the quadrupole interaction between the radio frequency electric field of the ion trap and the Rydberg electronic states. Such a %coupling is negligible in the customary ion traps but important for trapped \textit{Rydberg} ions due to their large electric dipole moment. Although the quadrupole interaction %oscillates with the trap frequency it cannot be neglected in the standard rotating-wave approximation. We investigate the effect of the quadrupole coupling by performing suitable %effective representation of the Hamiltonian.   For a single ion we show that in the effective picture the quadrupole interaction is replaced by rescaled laser intensities and %additional Stark shifts of the Rydberg levels. Moreover, we consider strong dipole-dipole interaction between the Rydberg ions in the presence of quadrupole coupling. In the %effective representation we observe reduction of the dipole-dipole coupling as well as an additional spin-spin interaction.

%\pacs{
%03.67.Lx, 03.67.Ac, 32.80.Qk, 42.50.Dv
%32.80.Xx, 	%Level crossing and optical pumping
%32.80.Qk, 	%Coherent control of atomic interactions with photons
%33.80.Be,   %Level crossing and optical pumping
%82.56.Jn 	%Pulse sequences in NMR
%42.50.Dv		%Quantum state engineering and measurements
%}

\maketitle

%\blue

%%%%%%%%%%%%%%%%%%%%%%%%%%%%%%%%%%%%%%%%%%%%%%%%%%%%%%%%%%%%%%%%%%%%%%%%%%%%%%%%%%%%%%%%%%%%%%%%%%%%%%%%%%%%%%%%%%%%%%%
\section{Introduction}
Ion trap system is one of the leading platforms in quantum information technologies \cite{Blatt2008,Haffner2012,Schneider2012}. The ability to control and read out the external and internal degrees of freedom of the trapped ions with high accuracy leads to experimental implementation of various entangled states \cite{Sackett2000,Haffner2005,Leibfried2005,Monz2011} and quantum gates with high fidelity \cite{Leibfried2003,Ballance2016,Gaebler2016}. However, the control of trapped ions becomes more difficult for multiple ions. Indeed, the phonon mode structure becomes too complicated for entanglement operations and the storage capacity is rendered limited. One way to overcome this limit is to use array of ion traps which store a small number of ions \cite{Kielpinski2002,Monroe2013,Weidt2016}. Another approach is based on using trapped \textit{Rydberg} ions. In this approach instead of using the common phonon mode for entanglement, the strong \textit{dipole-dipole interaction} may be used for implementation of entangled states and quantum gates \cite{Li2014}, as well as for quantum simulation \cite{Li2012}.

The strongly interacting Rydberg atoms offer a promising platform for quantum computation and simulation \cite{Saffman2010,Killian2007,Nguyen2018}. One hope that one may use the advantages of \textit{both} trapped ions (individual addressability, entanglement operations with small errors, etc.) \textit{and} the strong \textit{long} range interaction of Rydberg ions. However this novel system suffers from some disadvantages. For example, stray electric and magnetic fields due to the trap could alter the dipole moment of the Rydberg ion. Despite that, trapped Rydberg ions have been recently experimentally accomplished \cite{Feldker2015}. The Floquet sidebands due to quadrupole interaction as well as modification of the trapping potential due to the strong polarization of the $^{88}\text{Sr}^{+}$ Rydberg ion have been observed \cite{Higgins2017} .

\begin{figure}[tb]
  \includegraphics[width=0.45\textwidth]{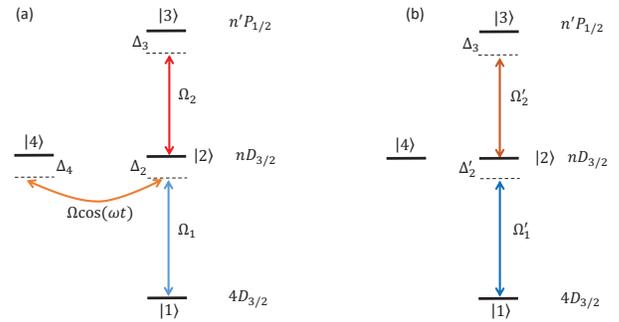}
  \caption{(a) Level scheme of the Rydberg $^{88}\text{Sr}^{+}$ ion. The levels $nD_{3/2}$ and $n^{\prime}P_{1/2}$ are Rydberg levels. Two laser fields are applied which drive the transitions $|1\rangle\leftrightarrow|2\rangle$ and $|2\rangle\leftrightarrow|3\rangle$ with Rabi frequencies $\Omega_{1,2}$ and detunings $\Delta_{2,3}$. The quadrupole field couples the levels $|2\rangle$ and $|4\rangle$ with peak Rabi frequency $\Omega$. This transition oscillates with radio trap frequency $\omega$. The level $|4\rangle$ has an energy shift $\Delta_{4}$. (b) Using the effective picture, the system is reduced to three state ladder system with rescaled Rabi frequencies $\Omega_{1,2}^{\prime}$ and detunings $\Delta_{2}^{\prime}$.}
  \label{FIG1}
\end{figure}

In Ref. \cite{Feldker2015,Bachor2016} the Rydberg levels are excited using a single-photon excitation with vacuum ultraviolet laser light at 122 nm. However, this is quite difficult to handle experimentally. Another experimental approach is to use $^{88}\text{Sr}^{+}$ Rydberg ions \cite{Higgins2017}. In that case the Rydberg ions are excited by two-photon transitions at 243 and 309 nm respectively. Though the $^{88}\text{Sr}^{+}$ ions are excited more easily to Rydberg levels, the $nD_{3/2}$ Rydberg states are coupled by the quadrupole field of the trap, see Fig. \ref{FIG1}. These undesirable transitions may transfer population out of the Rydberg state.

%%%%%%%%%%%%%%%%%%%%%%%%%%%%%%%%%%%%%%%%%%%%%%%%%%%%%%%%%%%%%%%%%%%%%%%%%%%%%%%%%%%%%%%%%%%%%%%%%%%%%%%%%%%%%%%%%%%%%%%%%%%%%%%%%%%%
%%%%%%%%%%%%%%%%%%%%%%%%%%%%%%%%%%%%%%%%%%%%%%%%%%%%%%%%%%%%%%%%%%%%%%%%%%%%%%%%%%%%%%%%%%%%%%%%%%%%%%%%%%%%%%%%%%%%%%%%%%%%%%%%%%%%
%%%%%%%%%%%%%%%%%%%%%%%%%%%%%%%%%%%%%%%%%%%%%%%%%%%%%%%%%%%%%%%%%%%%%%%%%%%%%%%%%%%%%%%%%%%%%%%%%%%%%%%%%%%%%%%%%%%%%%%%%%%%%%%%%%%%
%%%%%%%%%%%%%%%%%%%%%%%%%%%%%%%%%%%%%%%%%%%%%%%%%%%%%%%%%%%%%%%%%%%%%%%%%%%%%%%%%%%%%%%%%%%%%%%%%%%%%%%%%%%%%%%%%%%%%%%%%%%%%%%%%%%%

In this paper we consider the effect of the quadrupole coupling on the coherent dynamics of Rydberg trapped ions. Such an effect arises due to the composite nature of the Rydberg ion and causes undesired excitation of electronic transitions driven by the radio-frequency electric field of the Paul trap. We consider single Rydberg ion with one and two Rydberg state manifolds subject to the quadrupole coupling. We show that as long as the radio trap frequency $\omega$ of these quadrupole transitions is \textit{sufficiently large}, the negative effect is averaged and traced out. To see that, we perform a suitable unitary transformation and investigate the system into a different picture.  We show that the effect of  the quadrupole coupling is merely to \textit{rescale} the Rabi frequencies $\Omega_{i}$ which drive the transition between the Rydberg levels. The quandrupole interaction also induces an energy shift of the respective Rydberg levels. Moreover, we consider the strong dipole-dipole interaction between the Rydberg ions in the presence of quadrupole coupling. We show that the effect of the coupling is to reduce the strength of the dipole-dipole interaction. We also find that the quadrupole coupling induces residual dipole-dipole interaction which can be neglected only in the rotating-wave approximation.

The paper is organized as follows. In Section \ref{sec1} we provide discussion of the effect of the radio-frequency electric field on the electronic Rydberg transition. In Sec. \ref{sec2} we introduce the suitable unitary transformation which allows to treat the effect of the oscillating quadrupole coupling. In Sec. \ref{sec3} we discuss the single trap ion with one and two Rydberg state manifolds sensitive to the quadrupole coupling. In Sec. \ref{sec4} we consider two trapped Rydberg ions interacting via strong dipole-dipole coupling. Finally, in Sec. \ref{sec5} we summarize our findings.

%%%%%%%%%%%%%%%%%%%%%%%%%%%%%%%%%%%%%%%%%%%%%%%%%%%%%%%%%%%%%%%%%%%%%%%%%%%%%%%%%%%%%%%%%%%%%%%%%%%%%%%%%%%%%%%%%%%%%%%%%%%%%%%%%%%%
%%%%%%%%%%%%%%%%%%%%%%%%%%%%%%%%%%%%%%%%%%%%%%%%%%%%%%%%%%%%%%%%%%%%%%%%%%%%%%%%%%%%%%%%%%%%%%%%%%%%%%%%%%%%%%%%%%%%%%%%%%%%%%%%%%%%
\section{The level system of $^{88}\text{Sr}^{+}$ trapped ion}\label{sec1}
Our quantum system consists of a single trapped Rydberg ion. Although the method is applicable for any Rydberg ion we consider for concreteness Rydberg $^{88}\text{Sr}^{+}$ ion with the level structure shown on Fig. \ref{FIG1}. The Rabi frequency $\Omega_{1}$ drives the two-photon transition between the states $|1\rangle$ and $|2\rangle$. State $|2\rangle$ belongs to a Rydberg $nD_{3/2}$ manifold with detuning $\Delta_{2}$. We apply an additional laser field with Rabi frequency $\Omega_{2}$ which couples level $|2\rangle$ and level $|3\rangle$ with detuning $\Delta_{3}$. The latter is part of $n^{\prime}P_{1/2}$ manifold. The interaction Hamiltonian becomes ($\hbar=1$)
\begin{align}
\hat{H}_{0}&=\Delta_{2}|2\rangle\langle 2|+\Delta_{3}|3\rangle\langle 3|+\Delta_{4}|4\rangle\langle 4|\notag\\
&+\left(\Omega_{1}|1\rangle\langle 2|+\Omega_{2}|2\rangle\langle 3|+ \text{H.c.}\right).\label{HamInt1}
\end{align}

Let us consider the typical length scales of the trapped Rydberg ion. The external trapping frequency $\omega$ is of the order of MHz. To this frequency there corresponds a so called oscillator length $a_{\text{o}}$, which is roughly the \textit{localization length} of the ion around its \textit{equilibrium} position. For $\omega\sim \text{MHz}$ we have $a_{\text{o}}\sim 10 \;\text{nm}$. On the other hand, the size of the Rydberg \textit{orbit} $a_{\text{Ry}}$ is proportional to $n^{2}$, where $n$ is the principal quantum number. For Rydberg states $a_{\text{Ry}}\sim 100\;\text{nm}$. Thus, it follows that $a_{\text{Ry}}\gg a_{\text{o}}$. Therefore the Rydberg ion can no longer be considered as a point-like particle but rather as a \textit{composite} object \cite{Muller2008} and its internal structure \textit{must} be taken into account. Indeed, as shown in Refs. \cite{Muller2008,Kaler2011}, the electric field of the Paul trap may excite \textit{internal electronic} transitions which are no longer negligible contrary to the ordinary trapped ions.

The Paul trap electric field can be written as

\begin{equation}
\Phi(\textbf{r},t)=\alpha\cos{(\omega t)}(x^{2}-y^{2})-\beta\left(x^{2}+y^{2}-2z^{2}\right),
\end{equation}
where $\alpha$ and $\beta$ are electric field gradients and $\omega$ is the radio-frequency of the Paul trap \cite{Singer2010}. In the customary ion traps, this electric field does \textit{not} couple internal electronic states. The ion in that case can be considered as a point particle. However, in the case of Rydberg ions, it will couple electronic transitions. The coupling $\hat{H}_{\text{e}}$ of the above electric field is given by $\hat{H}_{\text{e}}=e\Phi(\textbf{r},t)$, where $e$ is the electronic charge. Generally, this quadrupole coupling \textit{cannot} couple (to first order) states in the manifold $nX_{J}$ for $J=1/2$ for any $X=S,P,D,...$. Such transitions are only allowed for $J>1/2$ due to selection rules. However, states in the manifold $nD_{J}$ ($J=3/2$ or $J=5/2$) are coupled even to first order by the quadrupole field. It turns out, that the time dependent interaction with the quadrupole is \cite{Higgins2017}
\begin{equation}
\hat{V}(t)=\hbar\Omega\cos{(\omega t)}\sum_{m_{J}=1/2}^{3/2}\{|nLJ(m_{J}-2)\rangle\langle nLJm_{J}|+\text{H.c.}\}\label{Quadrupole1},
\end{equation}
where $\Omega$ is the effective Rabi frequency for the quadruple coupling which oscillates with the trap frequency $\omega$. This transition may lead to a leak of population to an undesirable state $|4\rangle$ as is shown in Fig. \ref{FIG1}. Unfortunately rotating wave approximation is not applicable because the effective Rabi frequency $\Omega$ is comparable with the trap frequency $\omega$ \cite{Higgins2017}. In the next section we shall propose solution to this problem.

\section{General theory of the effective picture}\label{sec2}

First, let us rewrite Eq. ({\ref{Quadrupole1}}) for the $^{88}\text{Sr}^{+}$ ion,
\begin{equation}
\hat{V}(t)=\hat{v}e^{\i\omega t}+\hat{v}^{\dag}e^{-\i\omega t},
\end{equation}
where
\begin{equation}
\hat{v}=\frac{\Omega}{2}\left(|2\rangle\langle 4|+|4\rangle\langle 2|\right).\label{v-single}
\end{equation}
Including the quadrupole interaction the total Hamiltonian becomes
\begin{equation}
\hat{H}=\hat{H}_{0}+\hat{V}(t).\label{TotalHam}
\end{equation}
As we mentioned above the interaction $\hat{V}(t)$ may lead to leak of population out of Rydberg state $|2\rangle$ which spoils the single as well as the two qubit operators. In the following we perform a suitable unitary transformation. We shall designate this new quantum picture as an effective picture.

In order to derive the effective picture we perform a time dependent unitary transformation $\hat{U}(t)=e^{{\i}\hat{K}(t)}$ to the state vector $|\psi\rangle$ such that $|\tilde{\psi}\rangle=\hat{U}(t)|\psi\rangle$, where $\hat{K}(t)$ is an hermitian operator. Our goal is to choose $\hat{K}(t)$ such that the effective Hamiltonian $\hat{H}_{\text{eff}}=\hat{U}\hat{H}\hat{U}^{\dag}+i (\partial_{t}\hat{U})\hat{U}^{\dag}$ becomes a time-independent to \textit{any} desired order of $\omega^{-1}$. Method for averaging of the rapidly oscillating terms was proposed in \cite{James2007}, which however is not suitable for our case since it requires knowledge of the spectrum of $\hat{H}_{0}$. Wee derive $\hat{K}(t)$ following the method presented in \cite{Goldman2014,Rahav2003} (see the Supplement for an overview of the derivation). Here we simply state the result
\begin{eqnarray}
\hat{K}(t)=\omega^{-1}\hat{K}_{1}(t)+\omega^{-2}\hat{K}_{2}(t)+O(\omega^{-3}),\label{K(t)}
\end{eqnarray}
where
\begin{eqnarray}
\hat{K}_{1}(t)=2 \hat{v}\sin(\omega t),\quad \hat{K}_{2}(t)=-2i[\hat{v},\hat{H}_{0}]\cos(\omega t).\label{K(t)Corr}
\end{eqnarray}
We find that the effective Hamiltonian becomes
\begin{equation}
\hat{H}_{\rm eff}=\hat{H}_{0}+\omega^{-2}[[\hat{v},\hat{H}_{0}],\hat{v}]+O(\omega^{-4})\label{HEFFF},
\end{equation}
which is indeed time-independent to $O(\omega^{-4})$.

\section{Single Trapped Rydberg Ion}\label{sec3}

\subsection{Single manifold coupled by the quadrupole interaction}
%%%%%%%%%%%%%%%%%%%%%%%%%%%%%%%%%%%%%%%%%%%%%%%%%%%%%%%%%%%%%%%%%%%%%%%%%%%%%%%%%%%%%%%%%%%%%%%%%%%%%%%%%%%%%%%%%%%%%%%%%%%%%%%%%%
%%%%%%%%%%%%%%%%%%%%%%%%%%%%%%%%%%%%%%%%%%%%%%%%%%%%%%%%%%%%%%%%%%%%%%%%%%%%%%%%%%%%%%%%%%%%%%%%%%%%%%%%%%%%%%%%%%%%%%%%%%%%%%%%%%
\begin{figure}[tb]
  \includegraphics[width=0.45\textwidth]{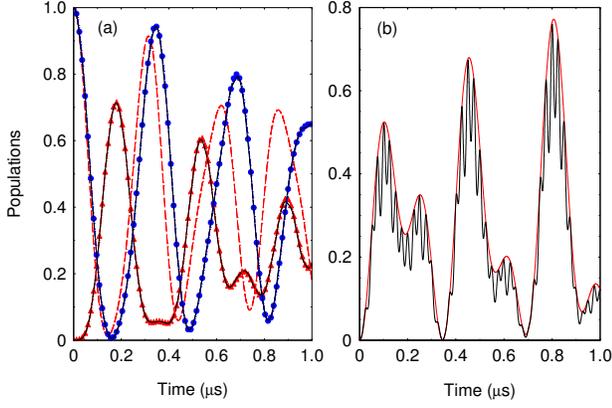}
  \caption{(a) Time evolution of the probabilities $P_{1}(t)$ and $P_{3}(t)$ for the four-level system. We compare the probabilities derived from the original Hamiltonian \eqref{TotalHam} (solid lines) and the effective Hamiltonian \eqref{Heff1} for $P_{1}(t)$ (blue dots) and $P_{3}(t)$ (red triangles). The red dashed line is the solution for $P_{1}(t)$ assuming rotating wave approximation. The parameters are set to $\Omega/2\pi=12$ MHz, $\omega/2\pi=20$ MHz, $\Omega_{i}/2\pi=2$ MHz, $\Delta_{2}/2\pi=\Delta_{3}/2\pi=2.0$ MHz, $\Delta_{4}/2\pi=1.0$ MHz. (b) Probability $P_{2}(t)$ (black solid line) compared with the effective solution (red line).}
  \label{FIG3}
\end{figure}
%%%%%%%%%%%%%%%%%%%%%%%%%%%%%%%%%%%%%%%%%%%%%%%%%%%%%%%%%%%%%%%%%%%%%%%%%%%%%%%%%%%%%%%%%%%%%%%%%%%%%%%%%%%%%%%%%%%%%%%%%%%%%%%%%%
%%%%%%%%%%%%%%%%%%%%%%%%%%%%%%%%%%%%%%%%%%%%%%%%%%%%%%%%%%%%%%%%%%%%%%%%%%%%%%%%%%%%%%%%%%%%%%%%%%%%%%%%%%%%%%%%%%%%%%%%%%%%%%%%%%
In this subsection we consider the single trapped Rydberg ion with \textit{one} Rydberg manifold coupled by the quadrupole coupling, see Fig. \ref{FIG1}.

Substituting Eqs. \eqref{HamInt1} and \eqref{v-single} into Eq. \eqref{HEFFF}, we obtain the following effective Hamiltonian,
\begin{align}
\hat{H}_{\text{eff}}&=\Delta_{2}^{\prime}|2\rangle\langle 2|+\Delta_{3}|3\rangle\langle 3|+\Delta_{4}^{\prime}|4\rangle\langle 4|\notag\\
&+\left(1-\frac{\Omega^{2}}{4\omega^{2}}\right)\left(\Omega_{1}|1\rangle\langle 2|+\Omega_{2}|2\rangle\langle 3|+\text{H.c.}\right).\label{Heff1}
\end{align}
Interestingly, we observe that the quadrupole interaction between states $|2\rangle$ and $|4\rangle$ is removed. However the new Rabi frequencies in the effective picture are rescaled (renormalized) with the same factor $\left(1-\Omega^{2}/(4\omega^{2})\right)$. Therefore in order to compensate the quadrupole interaction one needs to merely increase the laser intensities with the factor $\left(1-\Omega^{2}/(4\omega^{2})\right)^{-1}$. Additionally, we find that the quadrupole interaction caused an energy shift of the states $|2\rangle$ and $|4\rangle$ such that the laser detuning becomes $\Delta_{2}^{\prime}=\Delta_{2}\{1-\frac{\Omega^{2}}{2\omega^{2}}\left(1-\frac{\Delta_{4}}{\Delta_{2}}\right)\}$ and respectively
$\Delta_{4}^{\prime}=\Delta_{4}\{1-\frac{\Omega^{2}}{2\omega^{2}}\left(1-\frac{\Delta_{2}}{\Delta_{4}}\right)\}$.

\begin{figure}[tb]
  \includegraphics[width=0.45\textwidth]{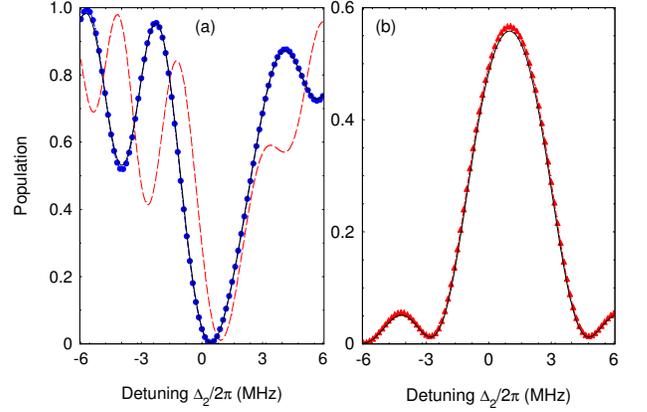}
  \caption{(a) Probability $P_{1}$ at time $t=0.5$ $\mu$s versus the laser detuning $\Delta_{2}$. The exact solution with the Hamiltonian \eqref{TotalHam} (solid lines) is compared with the solution with the effective Hamiltonian \eqref{Heff1} (blue circles). The red dashed line is the solution for $P_{1}$ assuming rotating wave approximation. (b) Same but for population $P_{3}$. Solid line is the exact result and the red triangle is the effective solution.}
  \label{d}
\end{figure}

In Fig. \ref{FIG3} we compare the exact dynamics governed by the full Hamiltonian (\ref{TotalHam}) and the effective Hamiltonian (\ref{Heff1}). As can be seen very good agreement is observed. We also show the effective dynamics which is obtained by standard rotating-wave approximation (RWA). As expected, RWA significantly deviates from the exact solution. This is due to the fact that $\Omega$ and $\omega$ are of the same order of magnitude, namely $\Omega=0.6\omega$. In Fig. \ref{FIG3}(b) we plot the population of the level $|2\rangle$ which is subject of the strong quadrupole interaction. Because of that the time evolution of the population contains fast and slow components where the latter can be described within the effective picture. Figure \ref{d} shows the frequency scan of the populations $P_{1,3}$ at fixed interaction time. The exact and the effective solutions are almost indiscernible.

After a lengthy calculation it can be shown that the next correction to the effective Hamiltonian Eq. (\ref{HEFFF}) is \textit{not} $O(\omega^{-3})$ but is $O(\omega^{-4})$. This explains why the agreement in Fig. \ref{FIG3} is quite accurate.

\subsection{Two Rydberg manifolds coupled by the quadrupole coupling}
%%%%%%%%%%%%%%%%%%%%%%%%%%%%%%%%%%%%%%%%%%%%%%%%%%%%%%%%%%%%%%%%%%%%%%%%%%%%%%%%%%%%%%%%%%%%%%%%%%%%%%%%%%%%%%%%%%%%%%%%%%%%%%%%%%
%%%%%%%%%%%%%%%%%%%%%%%%%%%%%%%%%%%%%%%%%%%%%%%%%%%%%%%%%%%%%%%%%%%%%%%%%%%%%%%%%%%%%%%%%%%%%%%%%%%%%%%%%%%%%%%%%%%%%%%%%%%%%%%%%%

\begin{figure}[tb]
  \includegraphics[width=1.0\columnwidth]{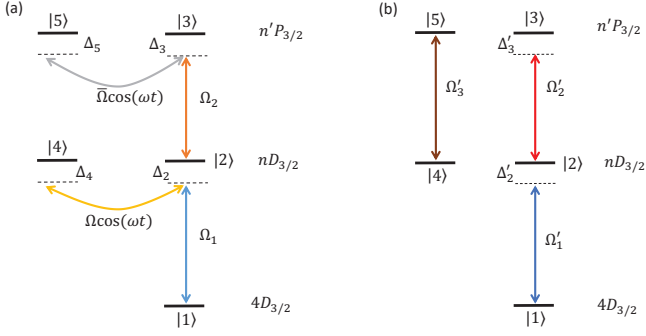}
  \caption{(a) Rydberg levels for a quadrupole coupling when \textit{both} Rydberg manifolds, $nD_{3/2}$ and $n^{\prime}P_{3/2}$ are coupled by the quadrupole trap field with Rabi frequencies $\Omega$ and $\bar{\Omega}$. Both quadrupole couplings oscillate with radio trap frequency $\omega$. (b) Effective quantum system is reduced into two \textit{uncoupled} systems. The first system consists of the levels $|4\rangle$ and $|5\rangle$ which are driven by Rabi frequency $\Omega_{3}^{\prime}=\frac{\Omega\bar{\Omega}\Omega_{2}}{2\omega^{2}}$.  The other system is formed by the states $|i\rangle$ $i=1,2,3$ in a ladder configuration driven by the rescaled Rabi frequencies $\Omega_{1}^{\prime}=\Omega_{1}\left(1-\frac{\Omega^{2}}{4\omega^{2}}\right)$ and $\Omega_{2}^{\prime}=\Omega_{2}\left(1-\frac{\Omega^{2}}{4\omega^{2}}-\frac{\bar{\Omega}^{2}}{4\omega^{2}}\right)$ and detunings $\Delta_{2}^{\prime}$, $\Delta_{3}^{\prime}$.}
  \label{FIG2}
\end{figure}

We extend the discussion by including higher angular momentum Rydberg states such as $n^{\prime}P_{3/2}$ states, see Fig. \ref{FIG2}(a). In that case the quadrupole Hamiltonian couples not only states $|2\rangle$ and $|4\rangle$ but also states $|3\rangle$ and $|5\rangle$. We shall show that in the effective picture the quadrupole coupling is again removed.

In this case the Hamiltonian is again of the type $\hat{H}=\hat{H}_{0}+\hat{V}(t)$. However, here
\begin{align}
\hat{H}_{0}=&\Delta_{2}|2\rangle\langle2|+\Delta_{3}|3\rangle\langle3|+\Delta_{4}|4\rangle\langle4|+\Delta_{5}|5\rangle\langle5|\notag\\
&+(\Omega_{1}|1\rangle\langle2|+\Omega_{2}|2\rangle\langle3|+{\rm H.c.})
\end{align}
and $\hat{V}(t)=\hat{v}e^{\i\omega t}+\text{H.c.}$, where $\hat{v}$ is given by
\begin{equation}
\hat{v}=\frac{\Omega}{2}|2\rangle\langle4|+\frac{\bar{\Omega}}{2}|3\rangle\langle5|+{\rm H.c.},
\end{equation}
with $\Omega$ and $\bar{\Omega}$ being the effective Rabi frequencies for the quadrupole interaction.

The expression Eq. (\ref{HEFFF}) for the effective Hamiltonian  as well as Eqs. (\ref{K(t)}) and (\ref{K(t)Corr}) for $\hat{K}$  remain valid. Thus, we obtain
\begin{equation}
\hat{H}_{\text{eff}}=\hat{H}_{1}+\hat{H_{2}}.\label{Heff2}
\end{equation}
Here
\begin{equation}
\hat{H}_{1}=\Delta_{4}^{\prime}|4\rangle\langle 4|+\Delta_{5}^{\prime}|5\rangle\langle 5|+\Omega_{3}^{\prime}\left(|4\rangle\langle 5|+|5\rangle\langle 4|\right),
\end{equation}
and
\begin{eqnarray}
\hat{H}_{2}&=&\Delta_{2}^{\prime}|2\rangle\langle2|+\Delta_{3}^{\prime}|3\rangle\langle3|+\Omega_{1}\left(1-\frac{\Omega^{2}}{4\omega^{2}}\right)(|1\rangle\langle2|\notag\\
&&+|2\rangle\langle1|)+\Omega_{2}\left(1-\frac{\Omega^{2}}{4\omega^{2}}-\frac{\bar{\Omega}^{2}}{4\omega^{2}}\right)\left(|2\rangle\langle3|+|3\rangle\langle2|\right).
\end{eqnarray}
This result means that the initial five-level coupled system is reduced to two uncoupled ladders, see Fig. \ref{FIG2}(b). The first ladder is a two level system consisting of states $|4\rangle$ and $|5\rangle$ driven by effective Rabi frequency $\Omega_{3}^{\prime}=\frac{\Omega\bar{\Omega}\Omega_{2}}{2\omega^{2}}$. This transition is caused by the virtual chain of transitions between the states $|4\rangle\leftrightarrow |2\rangle\leftrightarrow|3\rangle\leftrightarrow|5\rangle$. This explains why $\Omega_{3}^{\prime} \varpropto \Omega\Omega_{2}\bar{\Omega}$. Additionally, the quadrupole interaction causes energy shift of the level $|5\rangle$ such that we have $\Delta_{5}^{\prime}=\Delta_{5}\{1-\frac{\bar{\Omega}^{2}}{2\omega^{2}}\left(1-\frac{\Delta_{3}}{\Delta_{5}}\right)\}$. The second ladder consists of three states $|1\rangle$, $|2\rangle$ and $|3\rangle$. The effect of the quadruple interaction is to rescale the respective Rabi frequencies and detunings $\Delta_{2}^{\prime}$, $\Delta_{3}^{\prime}=\Delta_{3}\{1-\frac{\bar{\Omega}^{2}}{2\omega^{2}}\left(1-\frac{\Delta_{5}}{\Delta_{3}}\right)\}$ as is shown in Fig. \ref{FIG2}(b). As long as the initial population is in state $|1\rangle$, the population will remain in the second ladder, described by $\hat{H}_{2}$.
%%%%%%%%%%%%%%%%%%%%%%%%%%%%%%%%%%%%%%%%%%%%%%%%%%%%%%%%%%%%%%%%%%%%%%%%%%%%%%%%%%%%%%%%%%%%%%%%%%%%%%%%%%%%%%%%%%%%%%%%%%%%%%%%%%
%%%%%%%%%%%%%%%%%%%%%%%%%%%%%%%%%%%%%%%%%%%%%%%%%%%%%%%%%%%%%%%%%%%%%%%%%%%%%%%%%%%%%%%%%%%%%%%%%%%%%%%%%%%%%%%%%%%%%%%%%%%%%%%%%%
%%%%%%%%%%%%%%%%%%%%%%%%%%%%%%%%%%%%%%%%%%%%%%%%%%%%%%%%%%%%%%%%%%%%%%%%%%%%%%%%%%%%%%%%%%%%%%%%%%%%%%%%%%%%%%%%%%%%%%%%%%%%%%%%%%
%%%%%%%%%%%%%%%%%%%%%%%%%%%%%%%%%%%%%%%%%%%%%%%%%%%%%%%%%%%%%%%%%%%%%%%%%%%%%%%%%%%%%%%%%%%%%%%%%%%%%%%%%%%%%%%%%%%%%%%%%%%%%%%%%%

\begin{figure}[tb]
  \includegraphics[width=1.0\columnwidth]{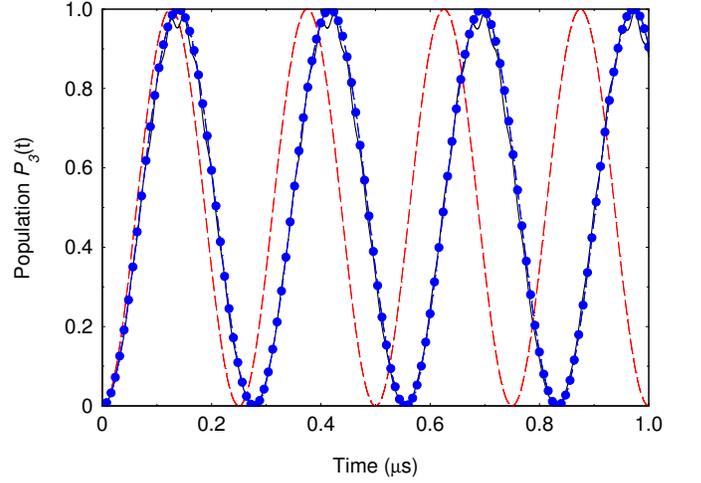}
  \caption{Time evolution of the probability $P_{3}(t)$. The Rydberg states $|2\rangle$ and $|3\rangle$ are coupled by quadrupole interaction with the states $|4\rangle$ and $|5\rangle$ with coupling strengths $\Omega/2\pi=12$ MHz and $\bar{\Omega}/2\pi=4$ MHz. The trap radio frequency is set to $\omega/2\pi=20$ MHz. The other parameters are $\Omega_{2}/2\pi=2$ MHz, $\Omega_{1}=0$, $\Delta_{i}=0$ ($i=2,3,4$). The solid line is the exact result and the dashed blue circles is the solution using the effective Hamiltonian (\ref{Heff2}). The red dashed line is the solution using rotating-wave approximation.}
  \label{FIG4}
\end{figure}

In Fig. \ref{FIG4} we show the resonance oscillations of the probability $P_{3}(t)$. We observe that the initial prepared population in state $|2\rangle$ exhibits Rabi oscillations where the exact solution is very closed to the effective picture. Although the quadrupole coupling between the states $|2\rangle$ and $|4\rangle$ is very strong and comparable with the radio trap frequency the corresponding probability is slightly affected.

\section{Two Rydberg ions interacting with dipole-dipole interaction}\label{sec4}

In this section, we extend the discussion including the dipole-dipole interaction. We consider an ion chain consisting of two Rydberg ions. The generalization for chain with $N$ ions is straightforward. The full Hamiltonian is quite complicated, see for example Ref. \cite{Muller2008}. However, under certain rather plausible approximations the Hamiltonian can be reduced to \cite{Muller2008}
\begin{equation}
\hat{H}=\hat{H}_{0}+\hat{V}(t)+\hat{H}_{\text{dd}}.\label{twoions}
\end{equation}
Here $\hat{H}_{0}$ is given by
\begin{align}
\hat{H}_{0}&=\sum_{j=1}^{2}\{\Delta_{2}|2_{j}\rangle\langle 2_{j}|+\Delta_{3}|3_{j}\rangle\langle 3_{j}|+\Delta_{4}|4_{j}\rangle\langle 4_{j}|\}\notag\\
&+\left(\Omega_{1}|1_{j}\rangle\langle 2_{j}|+\Omega_{2}|2_{j}\rangle\langle 3_{j}|+ \text{H.c.}\right)\}.
\end{align}
This is the single-ion Hamiltonian without the quadrupole interaction, see Fig. \ref{FIG1} and Eq. \eqref{HamInt1}.

The quadrupole interaction is again of the type $\hat{V}(t)=\hat{v}e^{\i\omega t}+\text{H.c.}$, where
\begin{equation}
\hat{v}=\frac{\Omega}{2}\left(|2_{1}\rangle\langle 4_{1}|+|2_{2}\rangle\langle 4_{2}|+\text{H.c.}\right).
\end{equation}
Lastly, the term $\hat{H}_{\text{dd}}$ is the dipole-dipole interaction. It is given by \cite{Muller2008}
\begin{equation}
\hat{H}_{\text{dd}}=\frac{\hat{d}_{1}^{(x)}\hat{d}_{2}^{(x)}+\hat{d}_{1}^{(y)}\hat{d}_{2}^{(y)}-2\hat{d}_{1}^{(z)}\hat{d}_{2}^{(z)}}{8\pi\epsilon_{0}|z_{0}^{(1)}-z_{0}^{(2)}|^{3}}.
\end{equation}
Here $\hat{d}_{j}^{(\alpha)},\;\alpha=x,y,z$ is the $\alpha$ component of the operator of the dipole moment for the $j$th ion, $\epsilon_{0}$ is the permittivity of the vacuum and $z_{0}^{(j)}$ is the equilibrium position of the $j$th ion along the $z$ axis. We can project this dipole-dipole interaction upon the basis states. Next we perform an optical RWA which is fulfilled as long as the Bohr transition frequencies of the Rydberg levels are much higher than the all Rabi frequencies, such that we obtain
\begin{equation}
\hat{H}_{\text{dd}}=\lambda\left(|2_{1}3_{2}\rangle\langle 3_{1}2_{2}| +|3_{1}2_{2}\rangle\langle 2_{1}3_{2}|\right) ,\label{Hdd}
\end{equation}
where
\begin{equation}
\lambda = \frac{|\langle 2|\hat{d}_{x}|3\rangle|^{2}+|\langle 2|\hat{d}_{y}|3\rangle|^{2}-2|\langle 2|\hat{d}_{z}|3\rangle|^{2}}{8\pi\epsilon_{0}|z_{0}^{(1)}-z_{0}^{(2)}|^{3}}.
\end{equation}
Here $\lambda$ is the strength of the Rydberg dipole-dipole interaction. Only the matrix elements of $\hat{d}_{\alpha},\;\alpha=x,y,z$ between \textit{Rydberg} states (state $|2\rangle$ and state $|3\rangle$) have been used, since the other matrix elements are negligible. The reason is that the overlap between the wave-function of the ground state $|1\rangle$ and a Rydberg wave-function is negligible. The dipole-dipole coupling resembles the XX Heisenberg spin-spin interaction. Indeed, setting $\Omega_{1}=0$ one can introduce the spin rising $\sigma^{+}_{j}=|3_{j}\rangle\langle2_{j}|$ and lowering $\sigma^{-}_{j}=|2_{j}\rangle\langle3_{j}|$ operators such that the dipole-dipole interaction can be rewritten as $\hat{H}_{\rm dd}=\lambda(\sigma^{x}_{1}\sigma^{x}_{2}+\sigma^{y}_{1}\sigma^{y}_{2})$, where $\sigma_{j}^{\alpha}$ are the Pauli matrices.
\begin{figure}[tb]
  \includegraphics[width=1.0\columnwidth]{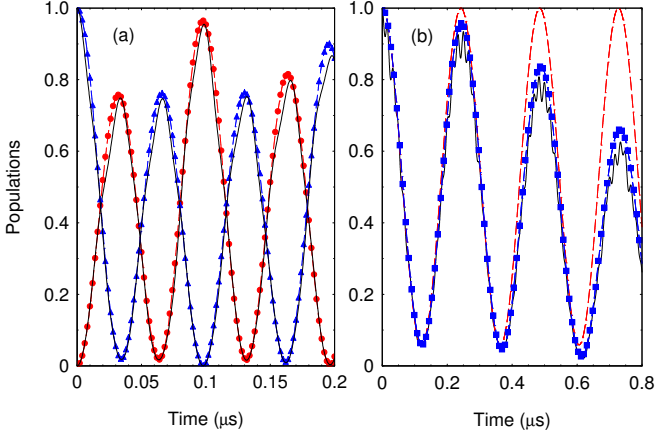}
  \caption{(a) Coherent exchange of spin excitation versus the interaction time. We compare the exact solution for the probabilities to observe states $|2_{1}3_{2}\rangle$ and $|3_{1}2_{2}\rangle$ (solid lines) with the effective Hamiltonian for $P_{23}(t)$ (blue triangles) and $P_{32}(t)$ (red circles). The parameters are set to $\Omega/2\pi=8.0$ MHz, $\omega/2\pi=30$ MHz, $\Delta_{2}/2\pi=1.0$ MHz, $\lambda/2\pi=7.0$ MHz, and $\Omega_{2}/2\pi=2.0$ MHz. (b) The same but initially the system is prepared in the state $|3_{1}4_{2}\rangle$. The solid line is the exact solution and the dashed blue squares is the solution with the effective Hamiltonian. The dashed line shows the probability $P_{34}(t)$ assuming rotating wave approximation.}
  \label{FIG5}
\end{figure}

Combining $\hat{\tilde{H}}_{0}=\hat{H}_{0}+\hat{H}_{\text{dd}}$, the total Hamiltonian becomes again of the type $\hat{H}=\hat{\tilde{H}}_{0}+\hat{V}(t)$. Therefore the expression (\ref{HEFFF}) for the effective Hamiltonian  as well as Eqs. (\ref{K(t)}) and (\ref{K(t)Corr}) for $\hat{K}$  remain valid. Using Eq. \eqref{HEFFF} the effective Hamiltonian becomes
\begin{align}
\hat{H}_{\text{eff}}&=\sum_{j=1}^{2}\{\Delta_{2}^{\prime}|2_{j}\rangle\langle 2_{j}|+\Delta_{3}^{\prime}|3_{j}\rangle\langle 3_{j}|+\Delta_{4}^{\prime}|4_{j}\rangle\langle 4_{j}|\}\notag\\
&+\left(1-\frac{\Omega^{2}}{4\omega^{2}}\right)(\Omega_{1}|1_{j}\rangle\langle 2_{j}|+\Omega_{2}|2_{j}\rangle\langle 3_{j}|+\text{H.c.})\}\notag\\
&+\lambda\left(1-\frac{\Omega^{2}}{2\omega^{2}}\right)\left(|2_{1}3_{2}\rangle\langle 3_{1}2_{2}|+|3_{1}2_{2}\rangle\langle 2_{1}3_{2}|\right)\notag\\
&+\frac{\lambda\Omega^{2}}{2\omega^{2}}\left(|3_{1}4_{2}\rangle\langle 4_{1}3_{2}|+|4_{1}3_{2}\rangle\langle 3_{1}4_{2}|\right).\label{Heff3}
\end{align}
In Fig. \ref{FIG5}, we compare the exact solution with Hamiltonian (\ref{twoions}) with the solution using the effective Hamiltonian (\ref{Heff3}). Due to the strong dipole-dipole interaction the system exhibits coherent exchange of spin excitations described by the XX Heisenberg spin model. The quadropule interaction leads to rescaling of the dipole-dipole coupling by the factor $\left(1-\Omega^{2}/(2\omega^{2})\right)$, i.e., $\lambda \rightarrow\lambda\left(1-\frac{\Omega^{2}}{2\omega^{2}}\right)$. The single ion Rabi frequencies are again renormalized with the same factor $\left(1-\Omega^{2}/(4\omega^{2})\right)$, i.e., $\Omega_{i}\rightarrow\Omega_{i}\left(1-\frac{\Omega^{2}}{4\omega^{2}}\right),\; i=1,2$. Additionally, the quadropule interaction induces residual dipole-dipole coupling between the states $|3_{i}4_{j}\rangle$ and $|4_{i}3_{j}\rangle$ described by the last term in (\ref{Heff3}). This coupling spoils the XX-type Heisenberg interaction between the Rydberg levels $|2_{i}3_{j}\rangle$ and $|3_{i}2_{j}\rangle$. In general, the residual dipole-dipole interaction can not be ignored except in the limit $\omega\gg\Omega$ where the RWA can be applied. Finally, we consider the dipole-dipole interaction between microwave dressed Rydberg ions. Such a dressing creates additional term in the dipole-dipole interaction (\ref{Hdd}) which couples the states $|2_{i}2_{j}\rangle\langle 2_{i}2_{j}|$ with coupling strengths $\mu$ (see the Supplement for more details). Because of that we find that the residual terms due to the quadrupole interaction are of order of $\mu(\Omega^{2}/2\omega^{2})$.

Note that the whole technique is valid so long as $\omega\gtrsim\Omega_{i},\; i=1,2$ \textit{as well as} $\omega\gtrsim\lambda$. The last condition $\lambda\lesssim\omega$ puts a lower limit on the frequency $\omega$. However the experimenter can increase $\omega$ above this limit. In addition, numerical simulations show that even for $\lambda=\omega/2$, the effective Hamiltonian remains quite correct. Therefore a long ranged dipole-dipole interaction of strength of $\sim 10\div 20\;\text{MHz}$ is still viable. In addition, by increasing the radio frequency $\omega$ more powerful interaction $\lambda$ can be used and the effective Hamiltonian is still applicable. For instance for $\omega = 2\pi\times 40\;\text{MHz}$ dipole-dipole interaction of the order of $20\;\text{MHz}$ can be achieved.

\section{Conclusion}\label{sec5}
In this paper we have shown that the quadrupole interaction which causes a reduction in the dipole moment in a trapped Rydberg ion can be dealt with by increasing the Rabi frequencies. To show that we have applied an unitary transformation. In this new picture, dubbed 'effective picture', see Fig. \ref{FIG1}, the Rabi frequencies are renormalized with the same factor $\left[1-\Omega^{2}/(4\omega^{2})\right]$. Therefore by increasing the laser intensities with the factor  $\left[1-\Omega^{2}/(4\omega^{2})\right]^{-1}$, the negative effect of the quadrupole interaction $\Omega\cos{\omega t}$ can be removed. In addition, we have extended the discussion, see Fig. \ref{FIG2}, when \textit{both} Rydberg manifolds are coupled by the quadrupole part of the trapping electric field. On the right side of Fig. \ref{FIG2} one observes that the effective five level system is \textit{decoupled} and the Rabi frequencies are altered by \textit{different} factors. Therefore even in that case, the negative effect of the quadrupole interaction can be removed. One merely has to rescale the laser intensities by different magnitudes. We have extended the discussion to an ion chain of two ions and we have shown that the Rabi frequencies are renormalized as well as the dipole-dipole coupling is modified. The latter is out of experimental control. However the reduction of the dipole-dipole coupling is only a few per cent for reasonable experimental parameters \cite{Higgins2017}, while the renormalization can be dealt with by increasing the Rabi frequency as was shown in the single ion case.

%%%%%%%%%%%%%%%%%%%%%%%%%%%%%%%%%%%%%%%%%%%%%%%%%%%%%%%%%%%%%%%%%%%%%%%%%%%%%%%%%%%%%%%%%%%%%%%%%%

\section*{Acknowledgments}
This work has been supported by the ERyQSenS, Bulgarian Science Fund Grant No. DO02/3.

\section*{Author Contributions}

L. S. S. developed the concept. The main calculations and numerical simulations are performed by L. S. S. and P. A. I. N. V. V. contributed to the analysis of the results. All authors wrote the manuscript.
\section*{Additional Information}
\textbf{Competing financial interests}: The authors declare no competing financial interests.

\end{document}